\documentclass[10pt,twocolumn,aps,prxquantum,superscriptaddress,showpacs,tightenlines,pdflatex,longbibliography]{revtex4-2}

\usepackage{amsthm}
\usepackage{gensymb}
\usepackage{amsmath}            
\usepackage{amssymb}            
\usepackage{mathtools}
\usepackage{graphicx}           
\usepackage{xcolor}
\usepackage{braket}
\usepackage{etoolbox}
\usepackage{epstopdf}
\usepackage{multirow}

\usepackage{xcolor}
\usepackage{bm}
\usepackage[normalem]{ulem}

\makeatletter
\let\cat@comma@active\@empty

\usepackage[colorlinks,citecolor=blue,urlcolor=blue,bookmarks=false,hypertexnames=true]{hyperref} 

\newcommand{\comment}[1]{}

\theoremstyle{plain}

\begin{document}
\title{RC\text{--}HEOM Hybrid Method for Non-Perturbative Open System Dynamics}

\author{Po-Rong Lai}
\affiliation{Department of Physics, National Cheng Kung University, Tainan 701, Taiwan}
\affiliation{Center for Quantum Frontiers of Research and Technology, NCKU, Tainan 701, Taiwan}

\author{Jhen-Dong Lin}
\affiliation{Department of Physics, National Cheng Kung University, Tainan 701, Taiwan}
\affiliation{Center for Quantum Frontiers of Research and Technology, NCKU, Tainan 701, Taiwan}

\author{Yi-Te Huang}
\affiliation{Department of Physics, National Cheng Kung University, Tainan 701, Taiwan}
\affiliation{Center for Quantum Frontiers of Research and Technology, NCKU, Tainan 701, Taiwan}

\author{Po-Chen Kuo}
\affiliation{Department of Physics, National Pingtung University, Pingtung 900, Taiwan }

\author{Neill Lambert}
\affiliation{RIKEN Center for Quantum Computing (RQC), Wakoshi, Saitama 351-0198, Japan}

\author{Yueh-Nan Chen}
\email{yuehnan@mail.ncku.edu.tw}
\affiliation{Department of Physics, National Cheng Kung University, Tainan 701, Taiwan}
\affiliation{Center for Quantum Frontiers of Research and Technology, NCKU, Tainan 701, Taiwan}
\affiliation{Physics Division, National Center for Theoretical Sciences, Taipei 106319, Taiwan}

\begin{abstract}

The Hierarchical equations of motion (HEOM) method is an important non-perturbative technique, allowing numerically exact treatment of open quantum systems with strong coupling and non-Markovian memory. However, its encoding of bath memory into auxiliary density operators often limits direct access to detailed bath information. In contrast, the reaction-coordinate (RC) mapping allows direct and transparent access to the dominant collective bath mode, but its perturbative and often Markovian treatment of the residual bath restricts its reliability. In this work, we introduce RC$\text{--}$HEOM, a hybrid method that unifies the strengths of both approaches by combining RC mapping with a fully non-perturbative HEOM description of the residual bath. RC$\text{--}$HEOM simultaneously retains exact non-Markovian memory and access to the RC mode, which enables analysis of system–RC information. Applying this method to the Anderson impurity models, we directly track the emergence of the Kondo singlet from the growth of the Kondo resonance and uncover a nontrivial RC-mediated coherence revival. These results demonstrate that RC$\text{--}$HEOM is a promising method for characterizing open quantum systems in regimes that are difficult to access with conventional master-equation methods.
\end{abstract}

\maketitle

\emph{Introduction}---
Open quantum systems research focuses on phenomena that occur when a system of interest interacts with a bath~\cite{breuer2002theory,Schaller2014,Rivas2012,Weiss2012}. For memoryless baths or weak system-bath coupling, Lindblad master equation approaches provide good explanations for quantum phenomena~\cite{Lindblad1976,Gorini1976,Manzano2020,Davies1974}. However, if the system-bath coupling is too strong or memory effects are too profound, then the bath's influence must be characterized by other approaches~\cite{Breuer2016,Rivas2014,devega2017}. Existing methods that allow partial access to the system-bath joint state and enable calculations of system-bath correlations include time-evolving density operator with orthogonal polynomials~\cite{Prior2010,Tamascelli2019} and numerical renormalization group~\cite{Bulla2008,Wilson1975}. These are chain-mapping approaches that map the bath into a chain of modes for calculations.

In recent years, development of reaction coordinate (RC) mapping in open quantum systems has shed light on an alternative approach towards obtaining bath information~\cite{Lambert2019,Strasberg2018,Ilessmith2014,Antosztrikacs2021,Shubrook2025,Latune2022,McConnell2022,Antosztrikacs2023}. By using a Bogoliubov transformation on the bath operators, the system-bath Hamiltonian is mapped exactly into another Hamiltonian. \textit{The Hamiltonian after RC mapping describes the system being coupled to the RC, which then couples to a residual bath}. Therefore, instead of solving the system dynamics when interacting with the bath, we can solve the system+RC dynamics when interacting with the residual bath. This allows us to obtain the system+RC joint state which provides information on the RC, the part of the bath that interacts with the system directly~\cite{Lambert2019,Strasberg2018,Ilessmith2014,Antosztrikacs2021,Shubrook2025,Latune2022,McConnell2022,Antosztrikacs2023}. 

Currently, most of the works use Lindblad master equations to obtain the dynamics of the system+RC joint state, but this method has limitations such as the need for a narrow bath spectral density~\cite{Lambert2019,Strasberg2018}. These limitations majorly come from the requirement of Born-Markov approximation when implementing Lindblad master equation~\cite{Manzano2020}. The RC mapping can be implemented multiple times for a more favorable residual bath spectral density, at a cost of computation power~\cite{Martinazzo2011}. As such, this approach is restricted to specific parameter regimes~\cite{Strasberg2018}.  \textit{To overcome these limitations, we consider a numerically exact treatment of the system+RC dynamics.}
\begin{figure}[!htbp]
\includegraphics[width=1\columnwidth]{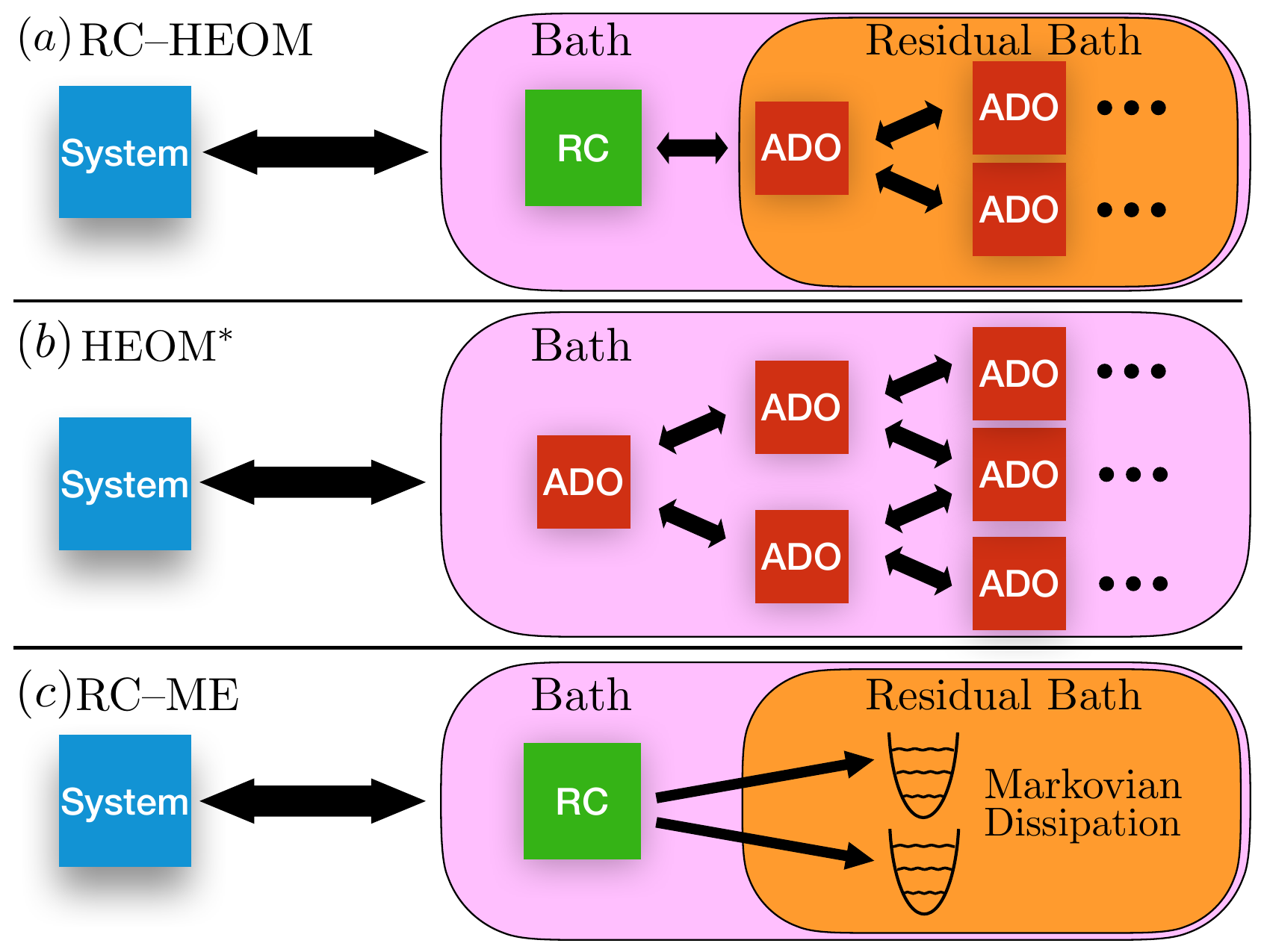}
\caption{Illustration of three methods used in open system dynamics calculations. (a) RC$\text{--}$HEOM. The system is coupled to the RC which couples to the ADOs of the residual bath. (b) HEOM*. System interactions with the bath is expressed as the system coupled to many ADOs. (c) RC$\text{--}$ME. The system is coupled to the RC which couples to the residual bath. System+RC interactions with the residaul bath are treated with the Lindblad master equation.}
\label{fig:method}
\end{figure}

In this letter, we develop a hybrid method that solves the system+RC dynamics exactly with the hierarchical equations of motion (HEOM~\cite{Tanimura1989,Tanimura1990,Tanimura2020,Cao2023,Li2012,Neill2023,Huang2023,Cirio2025}, where the high-order system-bath correlations are captured by introducing auxiliary density operators (ADOs).  Because \textit{the HEOM method is numerically exact} for systems coupled to Gaussian baths, this enables us to obtain the exact system+RC joint state.

To demonstrate the advantages of using this hybrid method (labeled as RC$\text{--}$HEOM), we consider two examples. For both examples, we compare RC$\text{--}$HEOM results with the traditional way of treating the residual bath using Lindblad master equation (labeled RC$\text{--}$ME), as well as using HEOM to solve the system dynamics without implementing RC mapping (labeled as HEOM*). We use HEOM* results as benchmarks to ensure RC$\text{--}$HEOM results are correct in contrast to the results using RC$\text{--}$ME. We use RC$\text{--}$HEOM to access RC state information to calculate system-RC correlations. We illustrate the three methods used in Fig.~\ref{fig:method}. We note the difference between the three methods in Table.~\ref{table:comparison}.

In the first example, we explore Kondo resonance of the single-impurity Anderson model in the Kondo regime~\cite{Anderson1961,Costi1994,GoldhaberGordon1998,Krishna-murthy1980,Hewson1993}. We find that RC$\text{--}$HEOM not only returns the correct density of states, but also allows us to calculate the singlet fraction~\cite{Modlawska2008}. We find that the singlet fraction increases as we lower the temperature, which indicates that the system+RC joint state is approaching the singlet state. In contrast, RC$\text{--}$ME fails to return the correct density of state to begin with. In the second example , we use RC$\text{--}$HEOM to explore the cause of bath mediated coherence revival in a two-impurity Anderson model~\cite{Andreani1993,Santoro1994,Eickhoff2018}. Unlike HEOM*, which only visualizes the coherence revival, \textit{RC$\text{--}$HEOM allows us to access system-bath coherence}. This enables us to understand how interference between effective system-bath coherence results in coherence revival of the two impurities.
\begin{table}[h]
  \centering
  \caption{Comparison of three methods.}\label{table:comparison}
  \begin{tabular}{c c c c}
    \hline\hline
     & RC\text{--}HEOM & HEOM* & RC\text{--}ME \\ \hline
    Numerically exact & \checkmark & \checkmark & \text{\sffamily X} \\[0.2em] 
    Computation speed & Slow & Slow & Fast  \\[0.4em]
    \shortstack{Access to \\Bath Information} & \shortstack{all system-RC\\information}  & \shortstack{certain\\observables} & \shortstack{all system-RC\\information} \\ \hline\hline
  \end{tabular}
\end{table}

\emph{RC$\text{--}$HEOM method}\label{sec:formalism}---
We start by detailing the RC$\text{--}$HEOM method. We begin with the RC mapping for a system linearly coupled to either a bosonic or fermionic bath. Then, we show how to utilize HEOM to characterize the system+RC interactions with the residual bath.

First, the total Hamiltonian before RC mapping (setting $\hbar=1$):
\begin{equation}\label{eq:general h}
\begin{aligned}
    &H_{\text{tot}}=H_{\text{sys}}+\sum_k\left(g_kc_k^{\dag}s+\text{h.c.}\right)+\sum_k\omega_kc_k^\dag c_k.
\end{aligned}
\end{equation}
Here, $H_{\text{sys}}$ is the Hamiltonian of the system mode $s$, $g_k$ is the coupling strength between the system and the $k$th bath mode, and $c_k$ is the $k$th bath mode with $\omega_k$ being its eigenenergy. The RC mapping can be described by a Bogoliubov transformation on the bath operators~\cite{Lambert2019,Strasberg2018,Ilessmith2014,Antosztrikacs2021,Shubrook2025,Latune2022,McConnell2022}. We choose a set of creation and annihilation operators ${C_k^{(\dag)}}$ via $C_k=\sum_l\Lambda_{kl}c_l$, where $\Lambda$ is a unitary matrix with $\Lambda\Lambda^\dag=\openone$. These operators obey commutation (anti-commutation) relations for a bosonic (fermionic) bath.  Further, the transformation satisfies two additional conditions:
\begin{equation}\label{eq:rc condition}
\begin{aligned}
    &\begin{cases}
        \lambda_0^*C_1 = \sum_k g_k^*c_k \\
        \sum_k\omega_k\Lambda_{lk}\Lambda_{mk}^*\equiv\delta_{lm}E_l~~\forall~m\neq1, l\neq1.
    \end{cases} 
\end{aligned}
\end{equation}
The first relation yields the system-RC coupling $\lambda_0$. The second relation generates the eigenenergy of the $l$th bath mode $E_l(l\geq2)$. After this RC mapping, the total Hamiltonian can be written as 
\begin{equation}\label{eq:general rch}
\begin{aligned}
    &H_{\text{tot}}= H_{\text{sys+RC}}+\sum_{l\neq1}\left(T_l^*C_1^{\dag}C_l+\text{h.c.}\right)+\sum_{l\neq1}E_lC_l^\dag C_l,
\end{aligned}
\end{equation}
where  $H_{\text{sys+RC}}=H_{\text{sys}}+\left(\lambda_0C_1^\dag s+\text{h.c.}\right)+E_1C_1^\dag C_1$. Here, we define $E_1\equiv\sum_k\omega_k|g_k|^2/|\lambda_0|^2$ to be the RC eigenenergy and $T_l\equiv\sum_m\omega_mg_m\Lambda_{lm}/\lambda_0$ is the RC coupling strength with the $l$th residual bath mode. The system is now coupled to the RC mode $C_1$ which then couples to the $l$ modes of the residual bath $C_l(l\geq 2)$. 

To obtain the system+RC joint state $\rho_{\text{sys+RC}}$, the RC$\text{--}$ME method considers Born-Markov approximation to obtain the following Lindblad master equation~\cite{Lambert2019,Strasberg2018,Ilessmith2014,Antosztrikacs2021}:
\begin{equation}
\begin{aligned}
    &\partial_t\rho_{\text{sys+RC}}(t)=-i[H_{\text{sys+RC}},\rho_{\text{sys+RC}}]+\hat{\mathcal{D}}[\rho_{\text{sys+RC}}]. 
\end{aligned}
\end{equation}
Here, $\hat{\mathcal{D}}[\cdot]\equiv\sum_k\gamma_k[L_k\cdot L_k^\dag-\{L_k^\dag L_k,\cdot\}/2]$ denotes the dissipator with decay rates $\gamma_k$ and jump operators $L_k=\ket{m}\bra{m}\hat{X}\ket{n}\bra{n}$, where $\hat{X}$ is the RC coupling operators $(C_1, C_1^\dag)$ and $\ket{m},\ket{n}$ are the eigenstates of the system+RC. Notably, this approach remains valid only when the interaction between the RC and the residual bath is sufficiently weak.

If we instead employ RC$\text{--}$HEOM, which treats the residual bath interactions without approximations, the reduced density matrix of the system+RC joint state at time $t$ can be written as
\begin{equation}\label{eq:dyson}
    \rho_{\text{sys+RC}}(t)=\hat{\mathcal{T}}e^{\mathcal{F}(t)} \rho_{\text{sys+RC}}(0),
\end{equation}
where $\hat{\mathcal{T}}$ is the time-ordering operator. Here, $\mathcal{F}(t)$ is the influence superoperator, which depends on the two-time correlation function $\mathcal{C}^\nu(t_1,t_2)$ of the residual bath and the RC mode $C_1$~\cite{Cirio2022,Lin2025,Huang2023}.

In order to numerically solve Eq.(\ref{eq:dyson}) with HEOM, one usually expresses the two-time correlation functions as a sum of exponents:
$\mathcal{C}^\nu(t_1,t_2)=\sum_{h=1}^N \eta_{\nu,h} e^{-\gamma_{\nu,h}(t_1-t_2)}$. This expression allows us to recursively differentiate the exponents in time to define local master equations for ADOs $\rho_{\text{sys+RC}}^{\textbf{q}}(t)$. Here, $\textbf{q}$ denotes a vector $[q_n,\cdots q_1]$ where each $q_i$ represents a multi-index ensemble $\{\nu,h\}$. These ADOs capture the system-bath correlations exactly. Therefore, solving their equations of motion results in the system+RC joint state, which is $\rho_{\text{sys+RC}}^{|\textbf{q}|=0}$. For $|\textbf{q}|>0$, the ADOs encode increasing orders of system-bath correlations. The RC$\text{--}$HEOM equations of motion can then be obtained:
\begin{equation}\label{eq:ado eom}
\begin{aligned}
    &\partial_t\rho_{\text{sys+RC}}^{\textbf{q}}(t)=-i[H_{\text{sys+RC}},\rho_{\text{sys+RC}}^{\textbf{q}}(t)] \\
    &-\sum_{w=1}^n\gamma_{q_w}\rho_{\text{sys+RC}}^{\textbf{q}}(t)-i\sum_{q'}\hat{\mathcal{A}}_{q'}\rho_{\text{sys+RC}}^{\textbf{q}^+}(t) \\
    &-i\sum_{w=1}^n\hat{\mathcal{B}}_{q_w}\rho_{\text{sys+RC}}^{\textbf{q}_w^-}(t),
\end{aligned}
\end{equation}
where we use the multi-index notations: $\textbf{q}^+=[q',q_n\cdots q_1]$ and $\textbf{q}_w^-=[q_n,\cdots,q_{w+1},q_{w-1},\cdots,q_1]$.
For fermions, we require $q'\notin\textbf{q}$ due to the Pauli exclusion principle. In addition, the superoperators $\hat{\mathcal{A}},\hat{\mathcal{B}}$ are used to describe system+RC interactions with the residual bath. They characterize how the $|\textbf{q}|$th level ADO ($|\textbf{q}|$ is the length of vector $\textbf{q}$) is coupled to the $|\textbf{q}^+|$th ADOs and the $|\textbf{q}^-_w|$th level ADOs. The mathematical details of the RC$\text{--}$HEOM formalism above and a bosonic example (since the two examples below work with fermionic systems) is shown in Supplemental Material~\cite{sup}. The numerical simulations in the following are performed using HierarchicalEOM.jl~\cite{Huang2023} and QuantumToolbox.jl~\cite{Mercurio2025}.

\emph{Example 1: Single impurity Anderson model}---
We explore the single impurity Anderson model (SIAM) using the RC$\text{--}$HEOM method. Starting from the total Hamiltonian of the SIAM (setting $\hbar=1$):
\begin{equation}\label{eq:siam H}
\begin{aligned}
    &H_{\text{tot}}=H_{\text{sys}}+\sum_{n=\uparrow,\downarrow}\sum_k\left(g_kc_{k,n}^{\dag}s_n\right. \\
    &\left.+g_k^*s_n^{\dag}c_{k,n}+\omega_kc_{k,n}^{\dag}c_{k,n}\right),
\end{aligned}
\end{equation}
where $H_{\text{sys}}=\epsilon(s_\uparrow^\dag s_\uparrow+ s_\downarrow^\dag s_\downarrow)+Us_\uparrow^\dag s_\uparrow s_\downarrow^\dag s_\downarrow$. Here, $\epsilon$ and $U$ are respectively the eigenenergy and repulsion energy of the impurity. Subscript $n=\uparrow,\downarrow$ denotes spin. We consider the Lorentzian bath spectral density:
\begin{equation}
J_0(\omega)=\frac{\Gamma W^2}{(\omega-\mu)^2+W^2}
\end{equation}
with strength $\Gamma$, chemical potential $\mu$, width $W$ and $\omega\in(-\infty,\infty)$. This allows us to derive the RC parameters and residual bath spectral density: $|\lambda_0|^2=\Gamma W/2, E_1=\mu, J_1(\omega)=2W$. We note that a flat residual spectral density is numerically difficult to implement using HEOM. Therefore, for the remainder of this letter, we employ an additional Lorentzian cutoff with sufficient width which returns the correct physics~\cite{sup}.

\begin{figure}[!htbp]
\includegraphics[width=1\columnwidth]{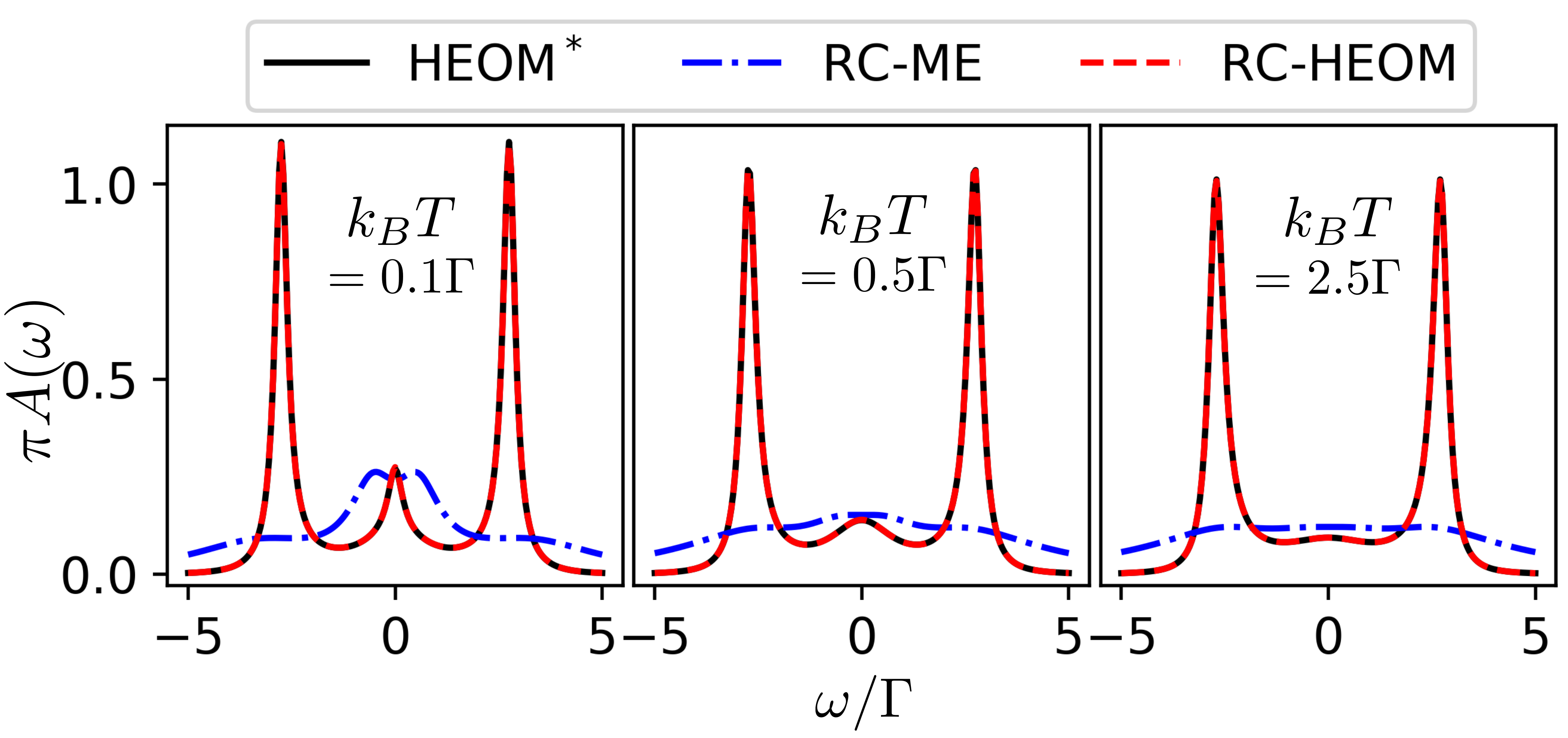}
\caption{The density of states $A(\omega)$ of the single impurity Anderson model at different temperatures. Lorentz bath parameters set to $W=1.25\Gamma,\mu=0$, and system parameters set to $U=3\pi\Gamma/2,\epsilon=-U/2$. Results using HEOM*, RC$\text{--}$ME, RC$\text{--}$HEOM are shown in black solid, blue dashdot, red dashed curves, respectively. Computation details in Supplemental Material~\cite{sup}.}
\label{fig:temp dos}
\end{figure}

The SIAM has long been used for the study of Kondo physics~\cite{Anderson1961,Bulla2008,Krishna-murthy1980,Hewson1993}. Particularly, both the formation of a Kondo singlet at zero temperature~\cite{Bulla2008,Wilson1975} and the temperature dependence of the Kondo resonance are well studied \cite{Nagaoka2002,Isidori2010}. Here, we take advantage of our RC$\text{--}$HEOM method to observe the singlet fraction $F$~\cite{Modlawska2008}, 
\begin{equation}
    F=\braket{\phi|\rho_{\text{sys+RC}}^\infty|\phi},
\end{equation}
which can be used to estimate how close our system+RC steady state $\rho_{\text{sys+RC}}^\infty$ is to the singlet state $\ket{\phi}=\frac{1}{\sqrt{2}}(\ket{\uparrow,\downarrow}_{\text{sys,RC}}-\ket{\downarrow,\uparrow}_{\text{sys,RC}})$. This measure indicates the local Kondo screening correlations captured by the system+RC joint state. As temperature decreases, Kondo screening increases~\cite{Shim2023,Tu2025}, thus, we expect the singlet fraction to increase as well. To ensure our calculations are correct, we first check the density of states~\cite{Cirio2023}
\begin{equation}
    A(\omega)=\frac{1}{\pi}\int dt e^{i\omega t}\braket{\{s_n(t),s_n^\dag(0)\}}
\end{equation}
to ensure that results obtained via RC$\text{--}$HEOM match the benchmark calculations from HEOM*. In Fig.~\ref{fig:temp dos}, 
we can see how the density of states obtained by HEOM* and RC$\text{--}$HEOM match up. The Kondo resonance increases as temperature drops as expected. We can also see that RC$\text{--}$ME is unable to produce the correct Kondo resonance, and the Hubbard resonance is not observable. This is due to our bath spectral density not being in the narrow structured regime opposite to the wide-band limit, where RC$\text{--}$ME is expected to be accurate~\cite{Strasberg2018}. 
We organize the results of the singlet fraction $F$ in Table.~\ref{table:singlet fraction} along with the Kondo resonance $A(\omega=0)$. As temperature drops, the Kondo resonance grows, and the system+RC steady state approaches the spin singlet $\ket{\phi}$. These results demonstrate RC$\text{--}$HEOM's capability to explore Kondo physics.
\begin{table}[h]
  \centering
  \renewcommand{\arraystretch}{1.5}
  \caption{Singlet fraction $F$ and Kondo resonance $A(\omega=0)$ for different temperatures.}\label{table:singlet fraction}
  \begin{tabular}{c @{\hspace{0.4cm}}c@{\hspace{0.4cm}} c@{\hspace{0.4cm}} c}
    \hline\hline
     $k_BT$ & $0.1\Gamma$ & $0.5\Gamma$ & $2.5\Gamma$ \\ \hline
     $\pi A(\omega=0)$ & 0.2741 & 0.1392 & 0.0935 \\ 
    $F$(RC$\text{--}$HEOM) & 0.2253 & 0.1754 & 0.0971 \\ \hline\hline
  \end{tabular}
\end{table}

\begin{figure}[!htbp]
\includegraphics[width=1\columnwidth]{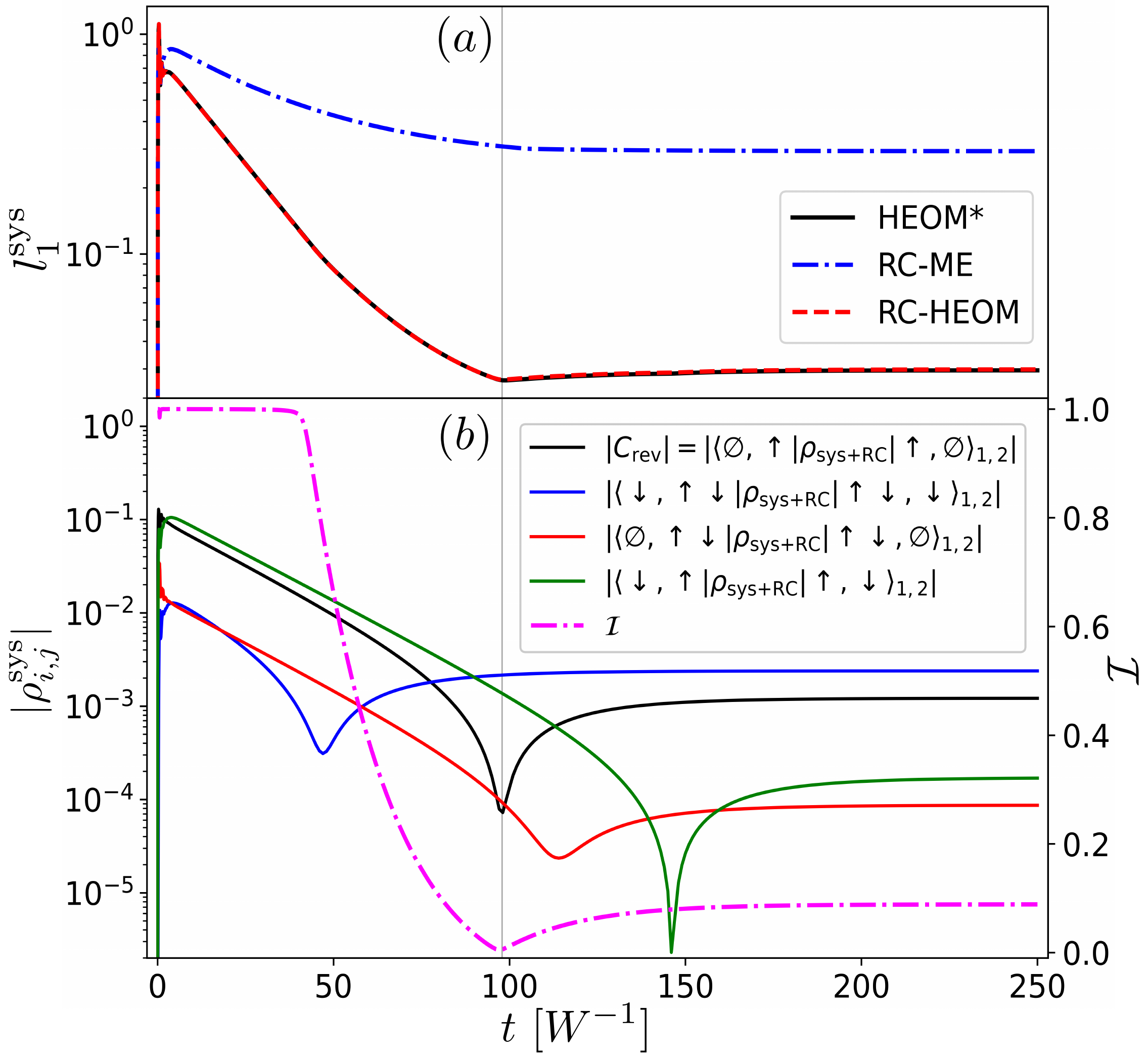}
\caption{Coherence revival of the two-impurity Anderson model. (a) $l_1$ norm of coherence for the two impurities: $l_1^{\text{sys}}$ with respect to time $t$. Results using HEOM*, RC$\text{--}$ME, RC$\text{--}$HEOM are shown in black solid, blue dashdot, red dashed curves, respectively.  Gray vertical line at $t'$ indicates the coherence revival time. (b) Individual coherence contributions $|\rho_{i,j}^{\text{sys}}|$ (that make up $l_1^{\text{sys}}$) which display revival are shown in black, blue, red, and green solid curves. Additionally, the interference factor $\mathcal{I}$ which measures the RC mediated path interference between the individual system+RC coherence that contribute to $|C_{\text{rev}}|$ (black solid curve) is shown as the magenta dashdot curve. Lorentz bath parameters are set to $\Gamma=20W, \mu=0, k_BT=5W$, and system parameters are set to$\epsilon_1=-2W, \epsilon_2=-W, U_1=U_2=10W$. Impurities are initialized in the vacuum state. RC is initialized in the thermal equilibrium state. Computation details are shown in Supplemental Material~\cite{sup}.}
\label{fig:tiam l1 and IF}
\end{figure}

\emph{Example 2: Bath mediated coherence revival in the two impurity Anderson model}--- The total Hamiltonian of the two-impurity Anderson model (TIAM) is written as:
\begin{equation}\label{eq:tiam H}
\begin{aligned}
    &H_{\text{tot}}=H_{\text{sys}}+\sum_{a=1,2}\sum_{n=\uparrow,\downarrow}\sum_k\left(g_kc_{k,n}^{\dag}s_{n,a}\right. \\
    &\left.+g_k^*s_{n,a}^{\dag}c_{k,n}\right)+\sum_{n=\uparrow,\downarrow}\sum_k\omega_kc_{k,n}^{\dag}c_{k,n},
\end{aligned}
\end{equation}
where $H_{\text{sys}}=\sum_{a=1,2}\epsilon_a(s_{\uparrow,a}^\dag s_{\uparrow,a}+ s_{\downarrow,a}^\dag s_{\downarrow,a})+U_as_{\uparrow,a}^\dag s_{\uparrow,a}s_{\downarrow,a}^\dag s_{\downarrow,a}$. Here, $\epsilon_a$ and $U_a$ are the eigenenergy and repulsion energy of impurity $a$, respectively. The bath spectral density is again Lorentzian. Notably, after RC mapping, the two impurities becomes coupled to a shared RC mode which then couples to the residual bath. Therefore, one can expect that the coherence between the two impurities could be mediated by the shared RC mode. To verify this intuition, we analyze the system coherence quantified by the $l_1$ norm~\cite{Yuan2020}:  $l_1^{\text{sys}}=\sum_{j\neq i}^{\text{dim}\{H_{\text{sys}}\}}|\rho_{i,j}^{\text{sys}}|$ with respect to time $t$. Here, $\rho_{i,j}^{\text{sys}}$ is the matrix element of the system density operator in the local occupation basis of the two impurties. 
We plot $l_1^{\text{sys}}$ in Fig.~\ref{fig:tiam l1 and IF}(a).

We find that the coherence dynamics using RC$\text{--}$HEOM agrees well with that of HEOM*, while the results using RC$\text{--}$ME does not. The key point of interest here is that HEOM* and RC$\text{--}$HEOM results both exhibit a sharp revival as indicated by a light gray vertical line at $Wt'=97.9$.

To identify the origin of this revival, we decompose $l_1^{\text{sys}}$ and analyze its individual compositions $|\rho_{i,j}^{\text{sys}}|$. We plot the compositions that display a revival in Fig.~\ref{fig:tiam l1 and IF}(b), as shown by the black, blue, red and green solid curves. We find that the black solid curve (along with its spin-down counterpart and conjugates that contribute identically) are the only coherence which exhibit a clear revival near $t'$. Because they connects states where each impurity hosts exactly one fermion
before and after the jump, we refer to them as one-fermion (OF)
coherence: $\bra{\varnothing,\uparrow}\rho_{\text{sys+RC}}\ket{\uparrow,\varnothing}_{1,2} ,~\bra{\uparrow,\varnothing}\rho_{\text{sys+RC}}\ket{\varnothing,\uparrow}_{1,2}$ and corresponding spin-down terms.

In order to check the significance of these OF coherence, we focus on  
\begin{equation}
    C_{\text{rev}}=\bra{\varnothing,\uparrow}\rho_{\text{sys+RC}}\ket{\uparrow,\varnothing}_{1,2},
\end{equation}
as shown by the black solid curve in Fig.~\ref{fig:tiam l1 and IF}(b). We observe its change in value after $t'$: $|C_{\text{rev}}(t_\infty)|-|C_{\text{rev}}(t')|\approx1.13\times10^{-3}$. Because the other OF coherence exhibit the same change over time as $|C_{\text{rev}}|
$, this means their total change is $\approx4.52\times10^{-3}$. If we compare this value with the change in the $l_1$ norm after $t'$: $l_1^{\text{sys}}(t_\infty)-l_1^{\text{sys}}(t')\approx2.97\times10^{-3}$, we can conclude that OF coherence are the dominant sources of coherence revival.

Next, we take advantage of RC$\text{--}$HEOM to access the RC's state information to understand how the RC mediates this revival. We decompose $C_{\text{rev}}$ into its components: $C_{\text{rev}}=C_{\text{rev}}^{\text{vac}}+C_{\text{rev}}^{\uparrow}+C_{\text{rev}}^{\downarrow}+C_{\text{rev}}^{\uparrow\downarrow}$, where each term corresponds to the RC being in the vacuum, spin-up, spin-down and doubly occupied state, respectively (e.g. $C_\text{rev}^{\text{vac}}=\bra{\varnothing,\uparrow,\varnothing}\rho_{\text{sys+RC}}\ket{\uparrow,\varnothing,\varnothing}_{1,2,RC}$). These four complex terms can interfere when we trace out the RC. Therefore, we can expect that $|C_{\text{rev}}|$ is also influenced by their interference. 

To capture this effect, we consider the interference factor $\mathcal{I}\in[0,1]$ defined as 
\begin{equation}
    \mathcal{I}\equiv\frac{|C_{\text{rev}}|}{|C_{\text{rev}}^{\text{vac}}|+|C_{\text{rev}}^{\uparrow}|+|C_{\text{rev}}^{\downarrow}|+|C_{\text{rev}}^{\uparrow\downarrow}|},
\end{equation}
where $\mathcal{I}=1$ indicates perfect constructive interference, whereas $\mathcal{I}=0$ indicates perfect destructive interference. We plot $\mathcal{I}$ against time in Fig.~\ref{fig:tiam l1 and IF}(b) as the magenta dashdot curve. We can observe that $|C_{\text{rev}}|$ rises and falls in tandem with $\mathcal{I}$, which is determined by how the four system+RC coherence: $C_{\text{rev}}^{\text{vac}},C_{\text{rev}}^{\uparrow},C_{\text{rev}}^{\downarrow},C_{\text{rev}}^{\uparrow\downarrow}$ interfere with each other.

We interpret the four system+RC coherence as the effective paths mediated by the RC to have a spin-up fermion jump from impurity 1 to impurity 2 $(C_{\text{rev}})$. Thus, the coherence revival originates from the overall interference of these paths changing from predominantly destructive to more constructive. The same mechanism applies to the other three OF coherence, so together they generate the revival of $l_1$ norm.

\emph{Conclusion}---We have developed RC–HEOM, a hybrid method that combines reaction-coordinate mapping with a non-perturbative HEOM treatment of the residual bath. This approach preserves exact non-Markovian memory while providing explicit access to the reaction-coordinate mode, thereby going beyond the limitations of both methods. We apply RC–HEOM to both a single and double Anderson impurity model and demonstrate that the method can track the formation of the Kondo singlet and reveal RC-mediated coherence revival, offering detailed insight into the interplay between the system and its environment. Overall, these results indicate that RC–HEOM is a useful and versatile tool for studying open quantum systems in regimes where conventional master-equation approaches are insufficient.

Given these features,  the proposed RC$\text{--}$HEOM method could be particularly useful for enhancing ongoing studies on strongly coupled or non-Markovian quantum thermodynamic devices and heat engines~\cite{Yu2014,GelbwaserKlimovsky2015,Katz2016,Newman2017,Ivander2022,Kaneyasu2023}, quantum transport~\cite{Restrepo2019,Wachtler2020,Antosztrikacs2022,Landi2022}, ultrastrong light–matter interactions~\cite{Cirio2016,Cirio2019,DeLiberato2017,Stassi2013,Beaudoin2011}, and non-Markovian heat and work statistics~\cite{Popovic2021,Nicolin2011,Kilgour2019}.

\section*{Acknowledgments}
This work is supported by the National Center for Theoretical Sciences and the National Science and Technology Council (NSTC), Taiwan, Grant No. NSTC 114-2112-M-006-015-MY3. P.-C. K. is supported by the National Science and Technology Council, Taiwan, under Grant No. NSTC 114-2112-M-153-004-MY3. N.~L.~is supported by MEXT KAKENHI Grant Numbers JP24H00816, JP24H00820. 


%

\end{document}


\title{Supplemental Material for\\
  ``RC\text{--}HEOM Hybrid Method for Non-perturbative Open System Dynamics''}

\author{Po-Rong Lai}
\affiliation{Department of Physics, National Cheng Kung University, Tainan 701, Taiwan}
\affiliation{Center for Quantum Frontiers of Research and Technology, NCKU, Tainan 701, Taiwan}

\author{Jhen-Dong Lin}
\affiliation{Department of Physics, National Cheng Kung University, Tainan 701, Taiwan}
\affiliation{Center for Quantum Frontiers of Research and Technology, NCKU, Tainan 701, Taiwan}

\author{Yi-Te Huang}
\affiliation{Department of Physics, National Cheng Kung University, Tainan 701, Taiwan}
\affiliation{Center for Quantum Frontiers of Research and Technology, NCKU, Tainan 701, Taiwan}

\author{Po-Chen Kuo}
\affiliation{Department of Physics, National Pingtung University, Pingtung 900, Taiwan }

\author{Neill Lambert}
\affiliation{RIKEN Center for Quantum Computing (RQC), Wakoshi, Saitama 351-0198, Japan}

\author{Yueh-Nan Chen}
\affiliation{Department of Physics, National Cheng Kung University, Tainan 701, Taiwan}
\affiliation{Center for Quantum Frontiers of Research and Technology, NCKU, Tainan 701, Taiwan}
\affiliation{Physics Division, National Center for Theoretical Sciences, Taipei 106319, Taiwan}

\maketitle

\section{Details of the RC$\text{--}$HEOM formalism}

We show the mathematical details of the reaction coordinate (RC) mapping, derive the RC parameters and residual bath spectral density and disclose more information on the reaction coordinate$\text{--}$hierarchical equations of motion (RC$\text{--}$HEOM).

\subsection{RC mapping}

We start from the total Hamiltonian (setting $\hbar=1$):
\begin{equation}\label{eq:general h}
\begin{aligned}
    &H_{\text{tot}}=H_{\text{sys}}+\sum_k\left(g_kc_k^{\dag}s+\text{h.c.}\right)+\sum_k\omega_kc_k^\dag c_k.
\end{aligned}
\end{equation}
Here, $H_{\text{sys}}$ is the Hamiltonian of the system with mode $s$, $g_k$ is the coupling strength between the system mode and $k$th bath mode, and $c_k$ is the $k$th bath mode with eigenenergy $\omega_k$. We choose the bath spectral density to be $J_0(\omega)=2\pi\sum_k|g_k|^2\delta(\omega-\omega_k)$. We begin the RC mapping by performing a Bogoliubov transformation on the bath operators. We choose a set of creation and annihilation operators ${C_k^{(\dag)}}$ via $C_k=\sum_l\Lambda_{kl}c_l$, where $\Lambda$ is a unitary matrix with $\Lambda\Lambda^\dag=\openone$. These operators obey commutation (anti-commutation) relations when we're working with a bosonic (fermionic) bath.  Specifically, we choose the operators that satisfy two conditions:
\begin{equation}\label{eq:rc condition}
\begin{aligned}
    &\begin{cases}
        \lambda_0^*C_1 = \sum_k g_k^*c_k \\
        \sum_k\omega_k\Lambda_{lk}\Lambda_{mk}^*\equiv\delta_{lm}E_l~\forall m\neq1, l\neq1.
    \end{cases} 
\end{aligned}
\end{equation}
The first condition generates 
\begin{equation}
    \sum_k\left(g_kc_k^\dag s+\text{h.c.}\right)=\lambda_0C_1^\dag s+\text{h.c.},
\end{equation}
where $\lambda_0$ is the system-RC coupling strength. By making use of commutation or anti-commutation relations, we obtain
\begin{equation}
\begin{aligned}
    &\lambda_0^*C_1\lambda_0C_1^\dag\mp\lambda_0C_1^\dag\lambda_0^*C_1=|\lambda_0|^2 =\sum_kg_k^*c_k\sum_lg_lc_l^\dag\mp\sum_lg_lc_l^\dag\sum_kg_k^*c_k \\
    &=\sum_{k,l}g_lg_k^*(c_kc_l^\dag\mp c_l^\dag c_k)=\sum_k|g_k|^2, \\
    &\text{where}~``-"~\text{stands for commutation relation},``+"~\text{stands for anti-commutation relation}.
\end{aligned}
\end{equation}
Thus, we connect the original system-bath coupling to the system-RC coupling.

By making use of the second condition in Eq.(\ref{eq:rc condition}), we obtain 
\begin{equation}
\begin{aligned}
    &\sum_k\omega_kc_k^\dag c_k=\sum_k\omega_k\sum_l\Lambda_{lk}C_l^\dag\sum_m\Lambda_{mk}^\dag C_m=\sum_{lm}C_l^\dag C_m\sum_k\omega_k\Lambda_{lk}\Lambda_{mk}^\dag.
\end{aligned}
\end{equation}
We can deduce that
\begin{equation}
\begin{aligned}
    &\begin{cases}
        C_1^{\dag}C_1\sum_k\omega_k\Lambda_{1k}\Lambda_{1k}^{\dag}=C_1^{\dag}C_1\sum_k\frac{\omega_k|g_k|^2}{|\lambda_0|^2}
        =E_1C_1^{\dag}C_1 & \text{if}~l=m=1,\\
        C_1^{\dag}\sum_mC_m\sum_k\omega_k\Lambda_{1k}\Lambda_{mk}^{\dag}
        =C_1^{\dag}\sum_mC_m\sum_k\omega_k\frac{g_k^*}{\lambda_0^*}\Lambda_{mk}^{\dag}
        =\sum_mT_m^*C_1^{\dag}C_m & \text{if}~l=1,m\neq1, \\

        \sum_lC_l^{\dag}C_1\sum_k\omega_k\Lambda_{lk}\Lambda_{1k}^{\dag}=\sum_lC_l^{\dag}C_1\sum_k\omega_k\frac{g_k}{\lambda_0}\Lambda_{lk}
        =\sum_lT_lC_l^{\dag}C_1 & \text{if}~l\neq1,m=1, \\ 

        \sum_{lk}C_l^{\dag}C_m\delta_{lm}E_l=\sum_lE_lC_l^{\dag}C_l & \text{if}~l\neq1,m\neq1,
    \end{cases}
\end{aligned}
\end{equation}
which implies $\sum_k\omega_kc_k^{\dag}c_k=E_1C_1^{\dag}C_1+\sum_mT_m^*C_1^{\dag}C_m 
    +\sum_lT_lC_l^{\dag}C_1+\sum_lE_lC_l^{\dag}C_l$, where we define $E_1\equiv\sum_k\frac{\omega_k|g_k|^2}{|\lambda_0|2}$ is the RC eigenenergy and $T_k\equiv\sum_m\frac{\omega_mg_m\Lambda_{km}}{\lambda_0}$ is the RC-residual bath coupling strength. Here, $E_l(l\geq2)$ is the eigenenergy of the $l$th residual bath mode. Under these settings the total Hamiltonian after RC mapping is: 
\begin{equation}\label{eq:general rch}
\begin{aligned}
    &H_{\text{tot}}= H_{\text{sys+RC}}+\sum_{l\neq1}\left(T_l^*C_1^{\dag}C_l+\text{h.c.}\right)+\sum_{l\neq1}E_lC_l^\dag C_l,
\end{aligned}
\end{equation}
where $H_{\text{sys+RC}}=H_{\text{sys}}+\left(\lambda_0C_1^\dag s+\text{h.c.}\right)+E_1C_1^\dag C_1$.

\subsection{RC parameters and residual bath spectral density}

Both the system-RC coupling strength and the RC eigenenergy as well as the residual bath spectral density can be expressed in terms of the bath spectral density $J_0(\omega)=2\pi\sum_k|g_k|^2\delta(\omega-\omega_k)$ (with cutoff frequencies $\omega_L$ and $\omega_R$, such that $\omega\in[\omega_L,\omega_R]$). The system-RC coupling and RC eigenenergy are:
\begin{equation}\label{eqapp:rc param}
\begin{aligned}
    &|\lambda_0|^2=\sum_k|g_k|^2=\int_{\omega_L}^{\omega_R}\frac{d\omega}{2\pi}J_0(\omega), \\
    &E_1=\sum_k\frac{\omega_k|g_k|^2}{|\lambda_0|^2}=\int_{\omega_L}^{\omega_R}\frac{d\omega}{2\pi}\frac{\omega J_0(\omega)}{|\lambda_0|^2}.
\end{aligned}
\end{equation}
To derive the residual bath spectral density $J_1(\omega)\equiv2\pi\sum_{l\neq1}|T_l|^2\delta(\omega-E_l)$, we start with Heisenberg's equations of motion for Eq.(\ref{eq:general h}):
\begin{equation}
\begin{aligned}
    &\dot{s}(t)=i[H_{\text{sys}}(t),s(t)]-i\sum_kg_k^*c_k(t)=-U_0(s)-i\sum_kg_k^*c_k(t), \\
    &\dot{c}_k(t)=-ig_ks(t)-i\omega_kc_k(t),
\end{aligned}
\end{equation}
where an operator $\mathcal{O}$ in the Heisenberg picture is $\mathcal{O}(t)=e^{-iH_{\text{tot}}t}\mathcal{O}e^{iH_{\text{tot}}t}(\hbar=1)$ and $U_0(s)=-i[H_{\text{sys}}(t),s(t)]$. Next, by performing Fourier transform $\hat{f}(z)=\int_{-\infty}^{\infty}e^{izt}f(t)dt$, we get
\begin{equation}
\begin{aligned}
    \begin{cases}
        -iz\hat{s}(z)=-\hat{U}_0-i\sum_kg_k^*\hat{c}_k(z) \\
        -iz\hat{c}_k(z)=-ig_k\hat{s}(z)-i\omega_k\hat{c}_k(z).
    \end{cases}
\end{aligned}
\end{equation}
Reorganizing terms around, we get
\begin{equation}\label{eqapp:w0}
\begin{aligned}
    &\hat{U}_0=iz\hat{s}(z)+i\frac{W_0(z)}{2}\hat{s}(z),
\end{aligned}
\end{equation}
where $W_0(z)\equiv2\sum_k\frac{|g_k|^2}{\omega_k-z}=\frac{1}{\pi}\int d\omega\frac{J_0(\omega)}{\omega-z}$. This gives us a $W_0$ function that links to $J_0$. Now we link it to $J_1$~\cite{Strasberg2018}. To do this, we write down Heisenberg's equations of motion for Eq.(\ref{eq:general rch}):
\begin{equation}
\begin{aligned}
    &\dot{s}(t)=-U_0(s)-i\lambda_0^*C_1(t), \\
    &\dot{C}_1(t)=-i\lambda_0s(t)-iE_1C_1(t)-i\sum_{l\neq1}T_l^*C_l(t), \\ 
    &\dot{C}_l(t)=-iT_lC_1(t)-iE_lC_l(t).
\end{aligned}
\end{equation}
After Fourier transform, we have
\begin{equation}
\begin{aligned}
    &-iz\hat{s}(z)=-\hat{U}_0-i\lambda_0^*\hat{C}_1(z), \\
    &-iz\hat{C}_1(z)=-i\lambda_0\hat{s}(z)-iE_1\hat{C}_1(z)-i\sum_{l\neq1}T_l^*\hat{C}_l(z), \\
    &-iz\hat{C}_l(z)=-iT_l\hat{C}_1(z)-iE_l\hat{C}_l(z).
\end{aligned}
\end{equation}
Reorganizing terms around, we get
\begin{equation}\label{eqapp:w1}
\begin{aligned}
    &\hat{U}_0=izs(z)+i\frac{|\lambda_0|^2\hat{s}(z)}{E_1-z-W_1(z)/2},
\end{aligned}
\end{equation}
where $W_1(z)\equiv2\sum_{l\neq1}\frac{|T_l|^2}{E_l-z}=\frac{1}{\pi}\int d\omega\frac{J_1(\omega)}{\omega-z}$. The LHS of Eq.(\ref{eqapp:w0}) and Eq.(\ref{eqapp:w1}) are the same, which gives us the connection between $W_0$ and $W_1$
\begin{equation}\label{eqapp:connection}
    \frac{W_0(z)}{2}=\frac{|\lambda_0|^2}{E_1-z-W_1(z)/2}.
\end{equation}
The next step is to use Sokhotski-Plemelj theorem~\cite{Haber2017,breuer2002theory}:
\begin{equation}
    \lim_{\Delta\rightarrow 0^+}\int_{\omega_L}^{\omega_R}\frac{f(\omega')}{\omega'-\omega-i\Delta}d\omega'=i\pi f(\omega)+\mathcal{P}\int_{\omega_L}^{\omega_R}\frac{f(\omega')}{\omega'-\omega}d\omega'.
\end{equation}
With $\mathcal{P}$ being the Cauchy principal, this gives us
\begin{equation}
\begin{aligned}
    &\Im[W_0(\omega)]=\Im[\lim_{\Delta\rightarrow0^+}W_0(z=\omega+i\Delta)]=J_0(\omega), \\
    &\Im[W_1(\omega)]=\Im[\lim_{\Delta\rightarrow0^+}W_1(z=\omega+i\Delta)]=J_1(\omega).
\end{aligned}
\end{equation}
Then, using Eq.(\ref{eqapp:connection}), we obtain
\begin{equation}\label{eqapp:residual SD}
    J_1(\omega)=\frac{4|\lambda_0|^2J_0(\omega)}{[\mathcal{P}\int_{\omega_L}^{\omega_R}\frac{1}{\pi}\frac{J_0(\omega')}{\omega'-\omega}d\omega']^2+J_0(\omega)^2}.
\end{equation}
Using Eq.(\ref{eqapp:rc param}) and Eq.(\ref{eqapp:residual SD}), we can derive that $|\lambda_0|^2=\Gamma W/2, E_1=\mu, J_1(\omega)=2W$ for a initial Lorentz spectral density $J_0(\omega)=\Gamma W^2/[(\omega-\mu)^2+W^2] (\omega\in(-\infty,\infty))$ in our main text.

\subsection{HEOM and equations of motion for ADOs}

For our RC$\text{--}$HEOM method, we treat the residual bath interaction with HEOM. Without approximations, the reduced density matrix of the system+RC joint state at time $t$ can be written in terms of the Dyson series~\cite{Cirio2022}:
\begin{equation}\label{eq:dyson}
\begin{aligned}
    &\rho_{\text{sys+RC}}(t)=\hat{\mathcal{T}}e^{\mathcal{F}(t)} \rho_{\text{sys+RC}}(0)
    =\hat{\mathcal{T}}\text{exp}\biggl\{-\int_0^t dt_1\int_0^{t_1}dt_2\hat{\mathcal{W}}(t_1,t_2)[\rho_{\text{sys+RC}}(0)]\biggr\},
\end{aligned}
\end{equation}
where $\hat{\mathcal{T}}$ is the time-ordering operator. Here, $\mathcal{F}(t)$ is the influence superoperator and a function of the two-time correlation function $\mathcal{CF}^\nu(t_1,t_2)$ and the RC mode $C_1$. The explicit form for $\hat{\mathcal{W}}$ for bosonic and fermionic baths are:
\begin{equation}
\begin{aligned}
    &\hat{\mathcal{W}}(t_1,t_2)[\cdot]=\sum_{p=\pm}\sum_{\nu=\pm}\biggl\{\mathcal{CF}^\nu(t_1,t_2)[C_1^{\bar{\nu}}(t_1), C_1^\nu(t_2)~\cdot~]_{-p}
    +\mathcal{CF}^\nu(t_2,t_1)[~\cdot~ C_1^{\bar{\nu}}(t_2),C_1^\nu(t_1)]_{-p}\biggr\}~\text{for fermions}, \\
    &\hat{\mathcal{W}}(t_1,t_2)[~\cdot~]=\sum_{\nu=\pm}\biggl\{\mathcal{CF}^\nu(t_1,t_2)[C_{1}^{\bar{\nu}},C_{1}^\nu(t_2)~\cdot~]_-
    -\mathcal{CF}^{\bar{\nu}}(t_2,t_1)[C_{1}^{\bar{\nu}}(t_1),~\cdot~ C_{1}^\nu(t_2)]_-\biggr\}~\text{for bosons},
\end{aligned}
\end{equation}
where $\nu$ denotes the presence ($\nu=+$) or absence ($\nu=-$) of Hermitian conjugation and $\bar{\nu}:=-\nu$. Here, $[\cdot,\cdot]_-$ and $[\cdot,\cdot]_+$ represent the commutator and anti-commutator to keep track of the parity $p$.

 Here, $\mathcal{CF}^\nu$ is the two-time correlation function for absorption ($\nu=+$) and emission ($\nu=-$) processes. We can express the correlation functions with the residual bath spectral density $J_1(\omega)$~\cite{Huang2023,Lin2025}
\begin{equation}\label{eqapp:correlation functions}
\begin{aligned}
    &\mathcal{CF}^{\nu=+}(t_1,t_2)=\int_{\omega_L}^{\omega_R}\frac{d\omega}{2\pi} J_1(\omega)\text{n}^{\text{Eq}}_{\text{FD}}(\omega)e^{i\omega(t_1-t_2)},~
    \mathcal{CF}^{\nu=-}(t_1,t_2)=\int_{\omega_L}^{\omega_R}\frac{d\omega}{2\pi} J_1(\omega)
    (1-\text{n}^{\text{Eq}}_{\text{FD}}(\omega))e^{-i\omega(t_1-t_2)}~\text{for fermions}, \\
    &\mathcal{CF}^{\nu=+}(t_1,t_2)=\int_{\omega_L}^{\omega_R}\frac{d\omega}{2\pi} J_1(\omega)\text{n}^{\text{Eq}}_{\text{BE}}(\omega)e^{i\omega(t_1-t_2)},~
    \mathcal{CF}^{\nu=-}(t_1,t_2)=\int_{\omega_L}^{\omega_R}\frac{d\omega}{2\pi} J_1(\omega)
    (1+\text{n}^{\text{Eq}}_{\text{BE}}(\omega))e^{-i\omega(t_1-t_2)}~\text{for bosons},
\end{aligned}
\end{equation}
where $\text{n}^{\text{Eq}}_{\text{FD}}(\omega)$ ($\text{n}^{\text{Eq}}_{\text{BE}}(\omega)$) is the Fermi-Dirac (Bose-Einstein) distribution. We can now see how the two-time correlation functions $\mathcal{CF}^\nu$ capture the bath effects.

In order to numerically solve Eq.(\ref{eq:dyson}) with HEOM, we usually express the two-time correlation functions as a sum of exponents:
$\mathcal{CF}^\nu(t_1,t_2)=\sum_{h=1}^N \eta_{\nu,h} e^{-\gamma_{\nu,h}(t_1-t_2)}$. This expression allows us to recursively differentiate the exponents in time to define local master equations for ADOs $\rho_{\text{sys+RC}}^{\textbf{q}}(t)$. Here, $\textbf{q}$ denotes a vector $[q_n,\cdots q_1]$, where each $q_i$ represents a multi-index ensemble $\{\nu,h\}$. These ADOs capture the system-bath correlations exactly, so solving their equation of motion results in the system+RC joint state, which is $\rho_{\text{sys+RC}}^{|\textbf{q}|=0}$. For $|\textbf{q}|>0$, the ADOs encode increasing orders of system-bath correlations. The RC$\text{--}$HEOM equations of motion are:
\begin{equation}\label{eq:ado eom}
\begin{aligned}
    &\partial_t\rho_{\text{sys+RC}}^{\textbf{q}}(t)=-i[H_{\text{sys+RC}},\rho_{\text{sys+RC}}^{\textbf{q}}(t)] 
    -\sum_{w=1}^n\gamma_{q_w}\rho_{\text{sys+RC}}^{\textbf{q}}(t)-i\sum_{q'}\hat{\mathcal{A}}_{q'}\rho_{\text{sys+RC}}^{\textbf{q}^+}(t) 
    -i\sum_{w=1}^n\hat{\mathcal{B}}_{q_w}\rho_{\text{sys+RC}}^{\textbf{q}_w^-}(t),
\end{aligned}
\end{equation}
where we use multi-index notations: $\textbf{q}^+=[q',q_n\cdots q_1],\textbf{q}_w^-=[q_n,\cdots,q_{w+1},q_{w-1},\cdots,q_1]$.
For fermions, we additionally require $q'\notin\textbf{q}$. Equation.(\ref{eq:ado eom}) uses superoperators $\hat{\mathcal{A}},\hat{\mathcal{B}}$ to describe system+RC interactions with the residual bath. They display how the $|\textbf{q}|$th level ADO ($|\textbf{q}|$ is the length of vector $\textbf{q}$) is coupled to the $|\textbf{q}^+|$th ADOs and the $|\textbf{q}^-_w|$th level ADOs. Their explicit forms are as follows:
\begin{equation}
\begin{aligned}
    &\hat{\mathcal{A}}_{q'}[\cdot]=C_1^{\bar{\nu}}[\cdot]-[\cdot]C_1^{\bar{\nu}},~
    \hat{\mathcal{B}}_{q_w}[\cdot]=(-1)^{n-w}\left(\eta_{\nu,h}C_1^\nu[\cdot]-(\eta_{\nu,h}^{\bar{\nu}})^*[\cdot]C_1^\nu\right)~\text{for fermions}, \\
    &\hat{\mathcal{A}}_{q'}[\cdot]=C_1^{\bar{\nu}}[\cdot]-[\cdot]C_1^{\bar{\nu}} 
    ,~\hat{\mathcal{B}}_{q_w}[\cdot]=\eta_{\nu,h}^\nu C_1^\nu[\cdot]-(\eta_{\nu,h}^{\bar{\nu}})^*[\cdot]C_1^\nu ~\text{for bosons}.
\end{aligned}
\end{equation}

\section{RC$\text{---}$HEOM computation details}
\subsection{Lorentz cutoff for residual bath spectral density}
When working with the single impurity Anderson model (SIAM) or two impurity Anderson model (TIAM) model coupled to a Lorentzian bath with the RC$\text{--}$HEOM method, we calculate the residual bath spectral density to be $J_1(\omega)=2W$, which is a constant. In this scenario, the two-time correlations functions would be
\begin{equation}
\begin{aligned}
    &\mathcal{CF}^{\nu=+}(t)=W\delta(t)-\frac{iWe^{i\mu t}}{\beta\sinh{\frac{t\pi}{\beta}}}~\text{and}~
    \mathcal{CF}^{\nu=-}(t)=W\delta(t)-\frac{iWe^{-i\mu t}}{\beta\sinh{\frac{t\pi}{\beta}}}.
\end{aligned}
\end{equation}
Unfortunately, there is no way of expressing the Dirac delta function $\delta(t)$ as a finite sum of exponents for us to use in HEOM. To circumvent this, instead of using $J_1(\omega)=2W$, we apply a Lorentzian cutoff so the residual bath spectral density looks like
\begin{equation}
    J_1'(\omega)=\frac{2W\cdot\Delta}{\omega^2+\Delta^2},~\text{where}~\Delta\gg W,
\end{equation}
which is effectively a very wide Lorentz spectral density as a replacement. This way, we can utilize the built in functions from HierarchicalEOM.jl to express the correlation functions as a sum of exponents and perform HEOM calculations~\cite{Huang2023}. 

To ensure that this replacement does not damage the integrity of our calculations, we compared key metrics with HEOM* results in our main text. We checked the density of states $A(\omega)=\frac{1}{\pi}\int dt e^{i\omega t}\langle\{s_n(t),s_n^\dag(0)\}\rangle$ in Example 1 and checked the coherence dynamics in Example 2. 

\subsection{HEOM computation parameters}
In Table \ref{table:table param}, we detail the computation parameters when using HEOM to generate the results in Example 1. The amount of memory and computation time needed to perform the calculations is decided by the size of the HEOM Liouvillian superoperator (HEOMLs) matrix. The HEOMLs matrix size is determined by "$\text{system dimension}^2\times\\~\text{number of ADOs}$"~\cite{Huang2023}. The number of auxiliary density operators (ADOs) is decided by the number of exponents used to fit the correlation function $N_{\text{exp}}$ and the maximum level of ADO we incorporate ($|\textbf{q}_{\text{max}}|$, known as the hierarchical tier or tier). The tables below show the $N_{\text{exp}}$ required to get a good fit of the correlation functions, whilst "tier" shows the number of hierarchical tiers necessary to reach convergence up to some small error. 

\begin{table}[!h]
  \centering
  \caption{HEOM computation parameters for Example 1 in the main text.}\label{table:table param}
  \begin{tabular}{c c c c c c c}
    \hline\hline
     & $k_BT$& system dimension & $N_{\text{exp}}$ & tier & ADOs & HEOMLs matrix size \\ \hline
    HEOM* & 0.2 & 4 & 4 & 3 & 1351 & 21,616 \\ 
    RC$\text{--}$HEOM & 0.2 & 16 & 4 & 4 & 6196 & 1,586,176\\ \hline
    HEOM* & 1 & 4 & 4 & 3 & 1351 & 21,616 \\ 
    RC$\text{--}$HEOM & 1 & 16 & 4 & 3 & 1351 & 345,856\\ \hline
    HEOM* & 5 & 4 & 2 & 2 & 79 & 1264 \\ 
    RC$\text{--}$HEOM & 5 & 16 & 2 & 2 & 79 & 20,224\\ \hline\hline
  \end{tabular}
\end{table}

In Table \ref{table:fig param}, we detail the computation parameters when using HEOM to generate the results in Example 2.

\begin{table}[!h]
  \centering
  \caption{HEOM computation parameters for Example 2 in the main text}\label{table:fig param}
  \begin{tabular}{c c c c c c}
    \hline\hline
     & system dimension & $N_{\text{exp}}$ & tier & ADOs & HEOMLs matrix size \\ \hline
    HEOM* & 16 & 2 & 5 & 1586 & 406,016 \\ 
    RC$\text{--}$HEOM & 64 & 6 & 2 & 407 & 1,667,072 \\ \hline\hline
  \end{tabular}
\end{table}

\section{Example of using RC$\text{--}$HEOM for bosonic systems}

We demonstrate how our RC$\text{--}$HEOM method can also work for bosonic systems. We demonstrate how we can obtain the correct system dynamics for the spin boson model with rotating wave approximation (RWA), i.e. the multi-mode Jaynes-Cummings model, when reaction coordinate$\text{--}$master equation (RC$\text{--}$ME) fails.

We start by performing the RC mapping on the model Hamiltonian:
\begin{equation}\label{eq:boson H}
    H_{\text{tot}} = H_{\text{sys}}+\sum_k\left(g_k\sigma_-b_k^{\dag}+g_k^*b_k\sigma_+\right)+\sum_k\omega_kb_k^{\dag}b_k
\end{equation}
Here, $H_{\text{sys}}=\frac{\omega_q}{2}\sigma_z+\frac{\Delta}{2}\sigma_x$ is the system Hamiltonian with qubit splitting $\omega_q$ and tunneling matrix element $\Delta$. Note that $\sigma_\pm$ are the system raising and lowering operators, $g_k$ is the system-bath coupling strength, $b_k$ is the $k$th mode of the bath with eigenenergy $\omega_k$. The system-bath correlations can be fully described by the spectral density $J_0(\omega)=2\pi\sum_k|g_k|^2\delta(\omega-\omega_k)$.

After RC mapping, the Hamiltonian is
\begin{equation}\label{eq:boson rch}
\begin{aligned}
    &H=H_{\text{sys}}+\lambda_0sB_1^{\dag}+\lambda_0^*B_1s^{\dag}+E_1B_1^{\dag}B_1
    +\sum_{l\neq1}\left(T_l^*B_1^{\dag}B_l+T_lB_l^{\dag}B_1\right) + \sum_{l\neq1}E_lB_l^{\dag}B_l,
\end{aligned}
\end{equation}
where $E_1$ is the RC eigenenergy, $T_l$ is the coupling strength between the RC mode $B_1$ and the $l$th bath mode $B_l(l\geq2)$, the latter with eigenenergy $E_l$. From Eq.(\ref{eq:boson rch}) we can see that the system is now coupled to the RC which then couples to the $l$ modes of the residual bath with spectral density $J_1(\omega)=2\pi\sum_{k\neq1}|T_k|^2\delta(\omega-E_k)$.

For our example, we choose the under-damped Brownian motion spectral density:
\begin{equation}
    J_0(\omega)=\frac{\gamma\lambda^2\omega}{(\omega^2-\omega_0^2)^2+\gamma^2\omega^2},
\end{equation}
where $\omega\in[0,\infty)$ and $\omega_0, \gamma, \lambda$ are the resonant frequency, width and strength of the bath, respectively. From which, we can derive the parameters of the Hamiltonian after reaction coordinate mapping:
\begin{equation}
\begin{aligned}
    &|\lambda_0|^2=\frac{\lambda^2}{\pi\sqrt{4\omega_0^2-\gamma^2}}\tan^{-1}{\frac{\sqrt{4\omega_0^2-\gamma^2}}{\gamma}}, \\
    &E_1=\frac{\lambda^2}{4|\lambda_0|^2}, \\
    &J_1(\omega)=\frac{4|\lambda_0|^2J_0}{(\frac{\omega_0^2-\omega^2}{\gamma\omega}J_0-\bar{p}(\omega))^2+J_0^2},
\end{aligned}
\end{equation}
where $\bar{p}(\omega)=\mathcal{P}\int_{-\infty}^{0}\frac{1}{\pi}\frac{J_0(\omega')}{\omega'-\omega}d\omega'$. Note that $\bar{p}(\omega)$ is solved numerically instead of the first term in the denominator in Eq.(\ref{eqapp:residual SD}). This is due to $\omega$ being positive in our example, so solving $\bar{p}(\omega)$ numerically is much easier because we don't integrate over any poles. 

Next, we implement HEOM to obtain the system+RC dynamics when interacting with the residual bath. We use the bosonic two-time correlation functions from Eq.(\ref{eqapp:correlation functions}).  The absorption and emission correlation functions must each be approximated as a sum of exponents to obtain the ADOs for HEOM calculation. To do so, we use the AAA algorithm to express $J_1(\omega), n_{\text{BE}}^{\text{Eq}}(\omega)$ in barycentric form~\cite{Nakatsuka2018}. Their poles are then used to express $\mathcal{CF}^\nu$ as a sum of exponents~\cite{Xu2022}. 

\begin{figure}[!htbp]
\includegraphics[width=0.5\columnwidth]{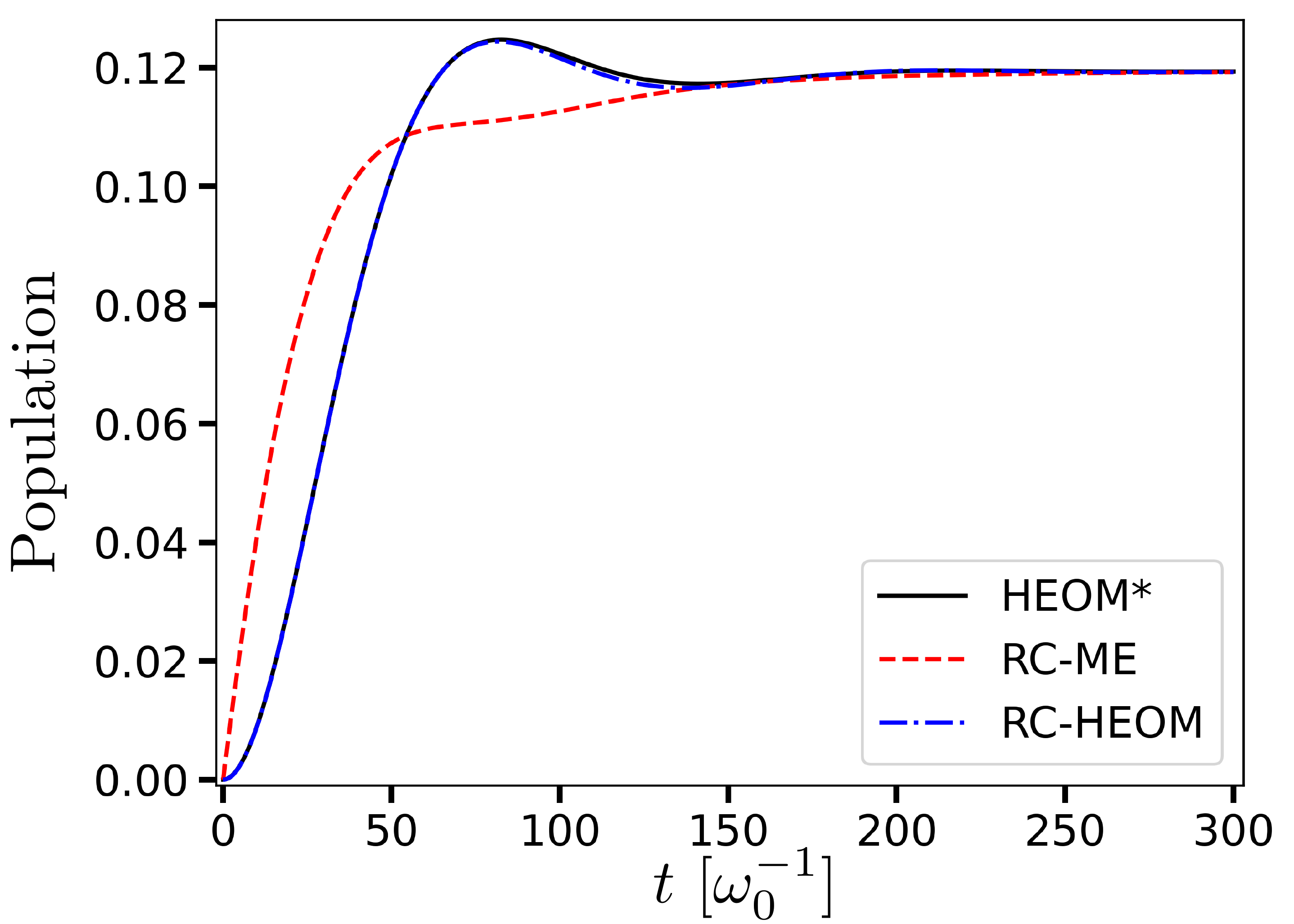}
\caption{System population with respect to time $t$. Black solid curve is the result using HEOM*, representing the correct dynamics. Red dashed curve is the result using RC$\text{--}$ME. Blue dashdot curve is the result using RC$\text{--}$HEOM. RC$\text{--}$HEOM recovers the correct short-time dynamics, whereas RC$\text{--}$ME does not, as seen in comparison with HEOM*.}
\label{fig:boson pop}
\end{figure}

We now look at the population dynamics when using RC$\text{--}$ME, HEOM* and RC$\text{--}$HEOM. We choose the bath spectral density parameters to be $\gamma=0.05\omega_0, \lambda=0.05\omega_0$ and set its temperature to be $k_BT=0.5\omega_0$. The system parameters are $\omega_q=\omega_0, \Delta=0$. We set the initial state of the system to be in its ground state, the RC in its thermal equilibrium state at temperature $T$. We plot the population dynamics against time $t$ (in units of $\omega_0^{-1}$) in Fig.\ref{fig:boson pop}. We find that when the population stabilizes, the results are the same for each method at around 0.119. However, RC$\text{--}$ME (red dashed) is unable to produce the correct short-time population dynamics. In contrast, using RC$\text{--}$HEOM (blue dashdot), the result is nearly identical to HEOM* (black solid). 

In Table \ref{table:boson param} below, we detail the computation parameters used to generate the results in Fig.\ref{fig:boson pop}. We write the system dimension when we use RC$\text{--}$HEOM as $2\times4$ to emphasize that the RC dimension here is 4. This means that it can have up to 3 bosons. This is the cutoff boson count that converges. 
\begin{table}[h]
  \centering
  \caption{HEOM computation parameters for bosonic example}\label{table:boson param}
  \begin{tabular}{c c c c c c}
    \hline\hline
     & system dimension & $N_{\text{exp}}$ & tier & ADOs & HEOMLs matrix size \\ \hline
    HEOM* & 2 & 88 & 3 & 121,485 & 485,940 \\ 
    RC$\text{--}$HEOM & 2$\times$4 & 256 & 2 & 33,153 & 2,121,792 \\ \hline\hline
  \end{tabular}
\end{table}

s


%